\documentclass[12pt,preprint]{aastex}
\def\lea{\mathrel{<\kern-1.0em\lower0.9ex\hbox{$\sim$}}}
\def\gea{\mathrel{>\kern-1.0em\lower0.9ex\hbox{$\sim$}}}

\slugcomment{Accepted for publication in ApJ Letters} 
\shorttitle{Faint Fuzzy Clusters in NGC 5195} \shortauthors{Hwang \& Lee}

\begin{document}

\title{Spatial Distribution of Faint Fuzzy Star Clusters in NGC 5195} 

\author{\sc Narae \ Hwang and Myung Gyoon \ Lee} 
\affil{Astronomy Program, SEES, Seoul National University, Seoul 151-742, Korea \\
	email: nhwang@astro.snu.ac.kr, mglee@astrog.snu.ac.kr}

\begin{abstract}
We present a study of a faint fuzzy star cluster system in the nearby SB0 galaxy NGC 5195 interacting with the famous spiral galaxy NGC 5194 (M51), 
based on $HST$ ACS $BVI$ mosaic images taken by the Hubble Heritage Team.
We have found about 50 faint fuzzy star clusters around NGC 5195 which are
 larger than typical globular clusters with effective radii $r_{eff}> 7$ pc and
red with $(V-I)>1.0$. 
They are mostly fainter than $M_V\approx -8.3$ mag.
From the comparison of $BVI$ photometry of these clusters with the simple stellar
population models, 
we find that they are as massive as $\approx 10^5 M_{\odot}$ and older than 1 Gyr. 
Strikingly, most of these clusters are found to be scattered in an elongated region almost perpendicular to the northern spiral arm of NGC 5194, and the center of the region is slightly
north of the NGC 5195 center,
while normal compact red clusters of NGC 5195 are located around the bright optical body of the host galaxy.
This is in contrast against the cases of NGC 1023 and NGC 3384 where spatial distribution of faint fuzzy clusters shows a ring structure around the host galaxy.
We suggest that at least some faint fuzzy clusters are experiencing tidal interactions 
with the companion galaxy NGC 5194 and must be associated with the tidal debris 
in the western halo of NGC 5195.

\end{abstract}


\section{INTRODUCTION}

Faint fuzzy star clusters are peculiar star cluster populations which are large ($r_{eff}> 7$ pc),
faint ($M_V>-8.0$), and red ($(V-I)>1.0$).
They are much larger than typical Galactic globular clusters, but are as red as old globular clusters.
These star clusters were introduced as a new family of star clusters 
by \citet{larb00} in their study of the nearby lenticular galaxy NGC 1023.
Although extensive searches over four early-type galaxies (three S0's and one E) were made by \citet{lar01},
these clusters were found only in two SB0 galaxies NGC 1023 (SB0) and NGC 3384 (SB0), 
and were not found in NGC 3115 (S0) and NGC 3379 (E1).

The main characteristics of these clusters derived from the imaging and spectroscopic studies \citep{larb00, larb02, bro02}
are summarized as follows:
(1) They are larger than normal old globular clusters. 
Their effective radii range from 7 to 15 pc, while the effective radii of typical globular clusters in 
our Galaxy are 2--3 pc;
(2) They are faint. Most of them are fainter than $M_V>-8$ mag;
(3) They are associated with disks of their host galaxies, showing a systematic rotation
similar to the rotation of the disk;
(4) Those in NGC 1023 are moderately metal-rich with [Fe/H]$=-0.58\pm0.24$ dex,
and are $\alpha$ element-enriched with [$\alpha$/Fe]$=0.3\sim 0.6$ dex; and 
(5) those in NGC 1023 are older than 7--8 Gyr;
(6) their origin is regarded to be associated with the interaction of host galaxies with their neighbors.

During this study, \citet{peng05} reported that 12 early-type galaxies in Virgo cluster harbor a significant number of diffuse star clusters, and that nine of them are S0s.
These clusters have low luminosities ($M_V>-8$), a broad spectrum of size ($3<r_h<30$ pc), and a broad
range of color ($0.3<(g-z)_0<1.9$), 
and they are predominantly red with $(g-z)_0>1.0$.
These clusters are mainly characterized by the $g$-band low surface brightness $\mu_g>20$ mag arcsec$^{-2}$.
The red population of these diffuse star clusters shares common features with faint fuzzy clusters, and
faint fuzzy clusters are considered to be a subset of the diffuse star clusters.

Recently, \cite{lee05} reported the discovery of a dozen of faint fuzzy clusters in NGC 5195,
which made this galaxy the third known SB0 galaxy inhabited by this rare family of star clusters.
However, due to the limited spatial coverage and filters ($VI$) of the $HST$ WFPC2 data they used, 
details of these clusters remain to be known: How many such clusters exist in NGC 5195?
How wide are they distributed? How old are they?

In the course of our study of star clusters in the M51 system, 
We have used $HST$ ACS mosaic images of NGC 5195 to get the answers to the above questions.
In this letter, we report a spatial distribution of these faint fuzzy star clusters in NGC 5195.

\section{DATA AND REDUCTION}

The data used in this study were obtained by the Hubble Heritage Team 
using $HST$ ACS with F435W(B), F555W(V), F814W(I), 
and F658N(H$\alpha$) filters through $HST$ observing program 10452 (P.I.: Steven V. W. Beckwith) 
to commemorate the $HST$'s 15th anniversary.
All the necessary data reduction processes were done by the STScI including multi-drizzling
for image combination before public release of the data.
Details of the data reduction are given in \citet{mut05} and their website (http://archive.stsci.edu/prepds/m51/).

We adopt a distance to NGC 5195 of $8.4\pm0.6$ Mpc ($(m-M)_0=29.62$), 
determined from the planetary nebula luminosity function in M51 \citep{fel97}. 
The corresponding linear scale is 40.7 parsecs per arcsecond.  The $HST$
ACS mosaic data have pixel sizes of $0.05\arcsec$ after the multi-drizzle process.

The foreground reddening toward NGC 5195 is low, $E(B-V)=0.035$
(corresponding extinctions are $A_B=0.150$, $A_V=0.115$, $A_I=0.067$ mag)
\citep{sch98}.
The total magnitudes and colors of NGC 5195 are
$B^T=10.45\pm0.07$ mag, $(B^T-V^T)=0.90\pm0.01$ \citep{dev91}.
For the adopted distance its absolute magnitudes are 
$M_B^T=-19.32$ mag and $M_V^T=-20.19$ mag, respectively.

\section{SELECTION OF STAR CLUSTERS}

At the distance of NGC 5195, typical stellar clusters will appear somewhat more
extended than stars in the $HST$ ACS images.
We made use of this information in radial
profiles of detected objects, plus other morphological criteria, in
order to separate clusters from stars and background galaxies.
Details of the source detection, photometry, and star cluster selection processes
are given in \citet{hwa05}, and only a brief description is presented here.

In order to identify star clusters using morphological information, we
utilized SEXTRACTOR \citep{ber96}, a program which builds catalogs of
objects from astronomical images and provides morphological
information as well. This routine 
provides a number of morphological
parameters which are very useful for separating star clusters from the
majority of stars and background galaxies.  

First, we identified all objects in the $V$ band images using SEXTRACTOR. 
A detection threshold of $4\sigma$ was used for finding objects in the images.
Secondly, we made a sample of star cluster candidates brighter than $V=24$ mag
using the morphological parameters provided by SEXTRACTOR. 
Finally we selected star clusters by visual inspection of the images of these candidates.
We divided the clusters into two classes: 
Class 1 for clusters with normal circular morphology, and Class 2 
for clusters with asymmetric morphology.
The number of star clusters we have found in the region of $6.8' \times 4.0'$ around NGC 5195
is 283 (246 for Class 1  and 37 for Class 2). 
We used only Class 1 clusters in this study.

We estimated the effective radii of these clusters using ISHAPE provided by \citet{lar99}.
For the point spread function (PSF) required for ISHAPE,
we constructed an empirical PSF
by carefully selecting isolated stars scattered evenly in the neighborhood of NGC 5195.
Variation of the PSF profile shape is found to be not significant, if any, around NGC 5195 field.

\section{RESULTS}

\begin{figure}
\plotone{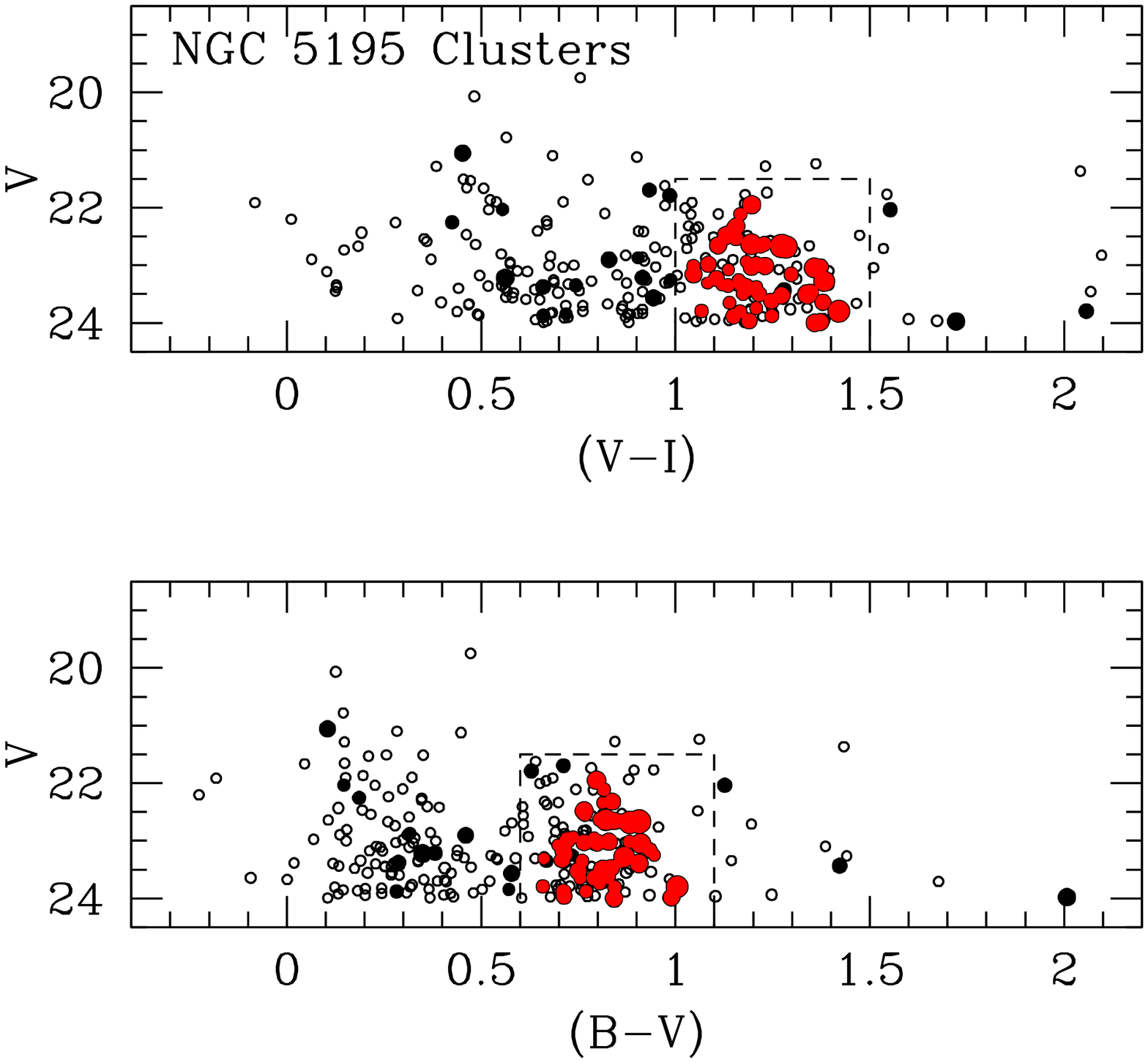}
\caption{
 $V-(V-I)$ and $V-(B-V)$ diagrams of all the measured star clusters in the $6.8' \times 4.0'$ field 
 around NGC 5195. 
 The size of the symbols is proportional to the effective radius $r_{eff}$ of each cluster.
 Clusters with $r_{eff} > 7$ pc are shown as filled symbols.
 Note the overdensity of large clusters with $r_{eff} > 7$ pc in $1.0<(V-I)<1.5$ and $0.6<(B-V)<1.1$
(dashed line boxes), which are selected as faint fuzzy clusters (red filled circles).
\label{fig1}}
\end{figure}

Fig.~\ref{fig1} shows the color-magnitude diagram (CMD) of 
all the measured star clusters in the $6.8' \times 4.0'$ field around NGC 5195.
One remarkable feature seen in Fig.~\ref{fig1} is that red clusters are, on average, quite larger than blue clusters.
We plot the effective radii vs. $(V-I)$ color of the clusters in Fig. 2.
The dashed line in Fig. 2 marks the effective radius $r_{eff}$ of 7 pc
used by \citet{larb00} 
to separate out extended, faint fuzzies from more typical, compact globular clusters.
Fig. 2 shows that
there are a significant number of large red clusters with $(V-I)>1.0$.  
The second feature in Fig.~\ref{fig1} is that extended clusters (marked by filled circles) 
are concentrated in a small range of colors.
We consider the star clusters with $r_{eff}>7$ pc and residing in both $1.0<(V-I)<1.5$ and $0.6<(B-V)<1.1$
color domain to be faint fuzzy clusters.
About 45 \% (49 out of 109) red clusters are found to be faint fuzzy star clusters and
all the faint fuzzy clusters discovered by \citet{lee05} were confirmed.
It should be noted that, however, our criteria to select faint fuzzy clusters are phenomenological,
just following \citet{larb00}.

We have derived age and mass of these clusters by fitting the $BVI$ photometry of the clusters
with the spectral energy distribution in the simple stellar population models of \citet{bc03}.
We adopted the solar metallicity ($Z=0.02$), following \citet{lee05}.
The derived ages of clusters range from 1.2 to 13.2 Gyr.
The range of ages will be  1.0 -- 3.4 Gyr for Z=0.05, and 2.4 -- 19.1 Gyr for Z=0.008, respectively.
The cluster masses are estimated to be $4.2 \times 10^4$ -- $3.4 \times 10^5 M_{\odot}$ for Z=0.02, 
$2.1 \times 10^4$ -- $2.4 \times 10^5 M_{\odot}$ for Z=0.05, and $6.0 \times 10^4$ -- $4.7 \times 10^5 M_{\odot}$ for Z=0.008.
This shows that these clusters are indeed older than 1 Gyr and as massive as $10^4$ -- $10^5 M_{\odot}$.

We display the spatial distribution of faint fuzzy star clusters of NGC 5195 in Fig.~\ref{fig3}:
filled circles for bright clusters with $V<23.3$ and open circles for faint clusters with $V \geq 23.3$.
The magnitude cut of $V=23.3$ is adopted to reduce any possible contamination due to background galaxies.
Of all the 17 galaxies with $V<24$ mag identified in the NGC 5195 field, 
five galaxies are found to be larger than
$r_{eff}=7$ pc and to belong to the color-magnitude parameter spaces of faint fuzzy clusters.
Among those galaxies, three galaxies with $V<23.3$ are found to show either 
very extended surface brightness profiles or evident non-circular morphologies.
This makes the distinction between galaxies and faint fuzzy clusters quite robust for 
the candidates brighter than $V=23.3$ mag.
Therefore, faint fuzzy clusters with $V<23.3$ are considered to be a safe sample with
negligible background galaxy contamination.

Several striking features are seen in Fig. 3.
First, faint fuzzy clusters are not distributed uniformly around the center of NGC 5195.
Instead, most of these clusters appear to be located in an elongated region almost perpendicular 
to the northern spiral arm of NGC 5194, while some clusters are scattered out of this area.  
On the other hand, compact red clusters in the lower panel
show quite a different distribution: 
most of them are distributed around the bright optical main body of NGC 5195.

Secondly, most blue star clusters with $(V-I)<0.7$ are composing a remarkably narrow sequence 
along the northern spiral arm of NGC 5194 passing through the eastern part of NGC 5195.
A few blue clusters that are located in the southern area of NGC 5195 
are considered to be members of NGC 5194.

Third, comparison with the distribution of planetary nebulae of M51 system \citet{dur03} shows some
common points with that of faint fuzzy clusters, especially in the western halo of NGC 5195.
\citet{dur03} pointed out that the complex kinematics of the planetary nebulae in this region 
may be due to tidal interaction in the M51 system.
Therefore it appears that some faint fuzzy clusters in this region may also belong to tidal debris.

Fourth, although the faint fuzzy cluster system of NGC 5195 shows an elongated structure,
it is not clear whether these clusters are distributed along a ring-like geometry,
as in the case of the faint fuzzy clusters in NGC 1023 \citep{larb00,bro02}.

Fifth, considering the theoretical evolutionary model fitting results that
the faint fuzzy clusters in NGC 5195 are older than 1 Gyr,
the difference in the spatial distribution between the faint fuzzy clusters and blue young clusters
can be attributed to the different formation epochs and environments:
the faint fuzzy clusters were formed around NGC 5195 before the interactions with NGC 5194,
while the young clusters are the results of the induced star formation triggered by
the interaction that took place about 100 $\sim$ 500 Myr ago \citep{sal00a, sal00b}.

\section{DISCUSSION AND CONCLUSION}

\citet{bro02} noted
that faint fuzzy clusters were found in only two late-type SB0 galaxies, NGC 1023 and NGC 3384,
among four early type galaxies investigated.
Various observations that found several features of interaction in NGC 1023 \citep{tul80, cap86} and NGC 3384 \citep{bus96, sil03} suggest that
the formation of faint fuzzy clusters in these galaxies are maybe related to the interaction of
their host galaxies with neighbors \citep{bro02}.
This is supported by theoretical studies \citep{fel02, fel05, bur05}.
\citet{peng05} noted that only two of the galaxies hosting diffuse star clusters have nearby neighbor galaxies.
We found from an inspection of the POSS maps and the data in NED\footnote{The NASA/IPAC Extragalactic Database, http://nedwww.ipac.caltech.edu.} of their sample 
that 40\% of 10 bright galaxies with $M_B<-18$ show hints of interaction 
(e.g., existence of neighbor galaxies, asymmetric outer halo) and 60\% of them are SB0 or SAB0.
Diffuse star clusters in bright galaxies ($M_B<-18$) are found either in
SB0s or in interacting galaxies. 
The only exception is VCC 1720 (SA), which happens to be the
faintest among the bright sample ($M_B=-18.9$). Therefore, major interaction of the host galaxy with its companions or the existence of a bar appears to be a key to the understanding of
the origin of faint fuzzy clusters and diffuse star clusters in these galaxies.

NGC 5195 is also a peculiar SB0 galaxy interacting with another barred spiral galaxy NGC 5194
which is two or three times more massive than NGC 5195 \citep{sch77}.
However, the estimated ages of the faint fuzzy clusters suggest that
it is not likely that these clusters were formed during the encounter with
NGC 5194.
This is in contradiction with the theoretical merging models proposed
for the formation of faint fuzzy clusters in NGC 1023 \citep{fel02}. 
This result indicates that the faint fuzzy clusters might have formed during the interaction
of galaxies much earlier than the age range investigated in the previous dynamical models or
they might have formed via some other mechanisms.

To date, NGC 1023 is the only galaxy which has been confirmed to have a rotating disk system
composed of faint fuzzy clusters \cite[]{bro02, bur05}.
\citet{peng05} found that diffuse star cluster populations in nine S0 galaxies in Virgo cluster
are both aligned with the host galaxy light and associated with galactic disks.
Nonetheless, the faint fuzzy clusters in NGC 5195 do not show any definite evidence in their
spatial distribution supporting their association with the optical disk of NGC 5195.
The elongated distribution of the faint fuzzy clusters looks different from
the bright optical body of NGC 5195 that is rather circular with ellipticity $e=0.21$ and inclined about $37^{\rm o}$
(north side to our direction) \cite[]{dev91}. 
However, faint stellar light is seen in the region where the faint fuzzy clusters are found.
In addition, the center of the elongated distribution of the faint fuzzy clusters is 
off about $1'$ northwest from the center of NGC 5195. 
Hence, it is not evident whether the faint fuzzy clusters belong to the disk of NGC 5195.

What if the elongated distribution should be the revelation of any physical structure unknown?
From HI observation, NGC 5195 was found to be nearly deficient in HI gas except for the gas associated
with the spiral arm of NGC 5194 touching its east side \cite[]{rot90}.
Only in the very center of NGC 5195, a high content of gases is detected from CO and HCN observation by \cite{koh02}.
Therefore we conclude that the elongated structure should be very old and devoid of gas.

It is also possible that the faint fuzzy clusters showing the elongated distribution
as a whole may belong to the debris of tidal interactions between NGC 5194 and NGC 5195. 
\citet{cha04} found 20 large globular clusters with $r_{eff}>7$ pc in the four $HST$ WFPC2 fields of 
NGC 5194, much more than in other spiral galaxies in their sample.
Nine of these clusters are similar to faint fuzzy clusters found in NGC 5195.
This indicates that faint fuzzy clusters might be dispersed not only over NGC 5195 but also over NGC 5194 through
the interactions.
It is noted, however, that about a dozen faint fuzzy clusters were found in one $HST$ WFPC2 field of NGC 5195
by \citet{lee05}, while only nine clusters were found in the four HST WFPC2 fields of NGC 5194.

Recently, \citet{bur05} suggested a scenario that 
the passage of a small companion galaxy through or near the center of a disk galaxy
could result in the annular distribution of faint fuzzy clusters in NGC 1023.
In M51 system, the mass of NGC 5195 is 1/2 to 1/3 of that of NGC 5194 
and the faint fuzzy clusters are not in a well-defined annular but in an elongated distribution,
which is quite a different case from NGC 1023.
Therefore \citet{bur05}'s scenario may not be applied to the case of NGC 5195.

In summary, we present a study of faint fuzzy clusters in NGC 5195 based on the analysis
of the HST ACS $BVI$ mosaic images.
We found about 50 such clusters in the $6.8' \times 4.0'$ field around NGC 5195.
Most of the faint fuzzy clusters appear to be distributed over an elongated region whose
center is about one arcmin off from the center of NGC 5195.
while compact red clusters are located around NGC 5195. 
Some faint fuzzy clusters are at least regarded as parts of
tidal debris produced by the interactions between NGC 5194 and NGC 5195.
Spectroscopic studies are needed to investigate the age, metallicity, and kinematics of these faint fuzzy clusters and to understand their origin.

\acknowledgements

N.H. and M.G.L. acknowledge the support of the BK21 program of Korean Government and
thank Yanbin Yang for providing the SED fitting programs.
M.G.L. is supported in part by the KOSEF grant (R01-2004-000-10490-0).
The authors are grateful to the anonymous referee for very helpful comments that improved our original manuscript.

\begin{figure}
\plotone{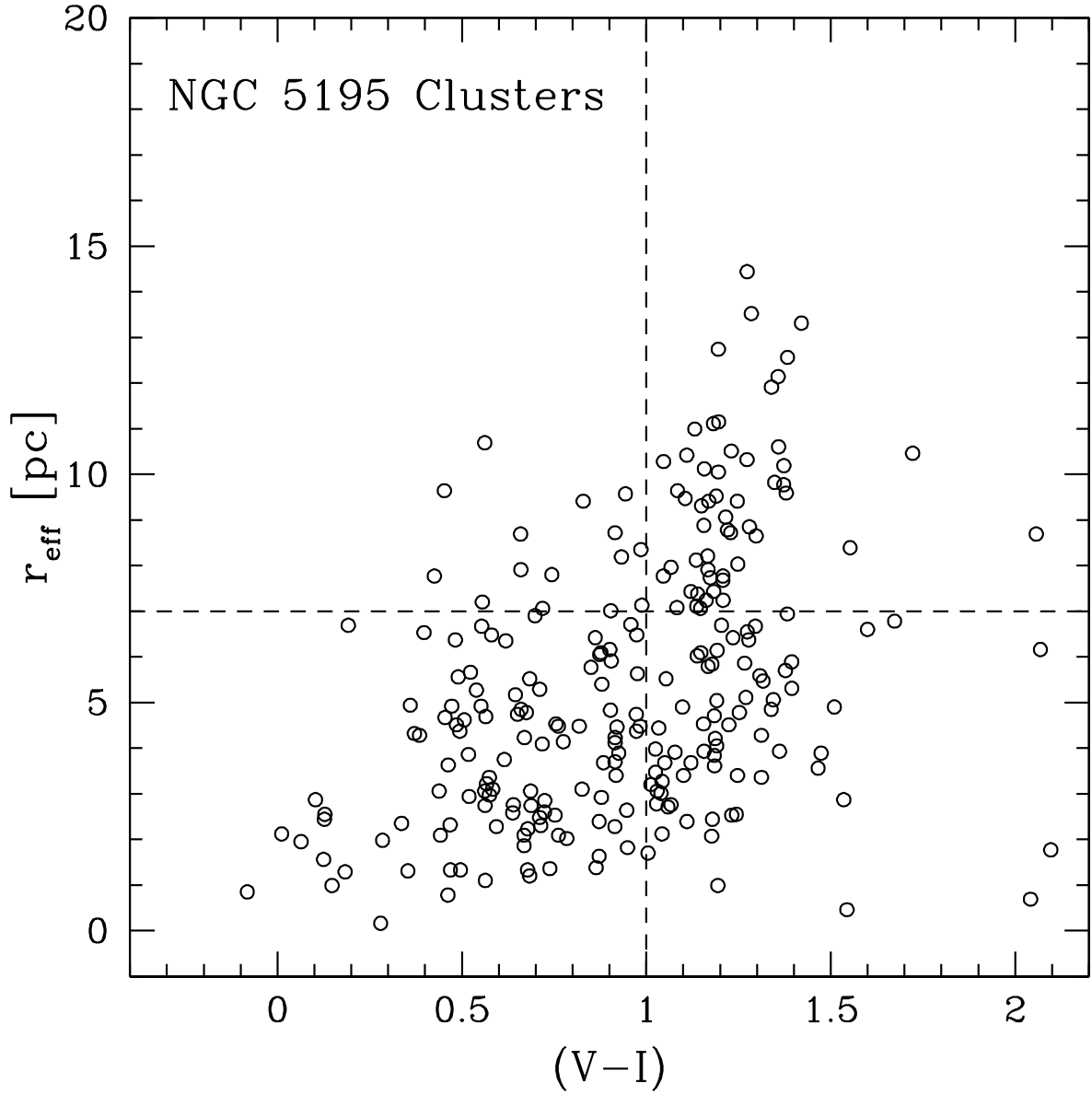}
\caption{
Effective radii $r_{eff}$ (pc) vs. $(V-I)$ of star clusters 
in the field of NGC 5195.
Dashed line marks $r_{eff}=7$ pc used as a boundary for selecting faint fuzzy clusters. 
\label{fig2}}
\end{figure}

\begin{figure}
\plotone{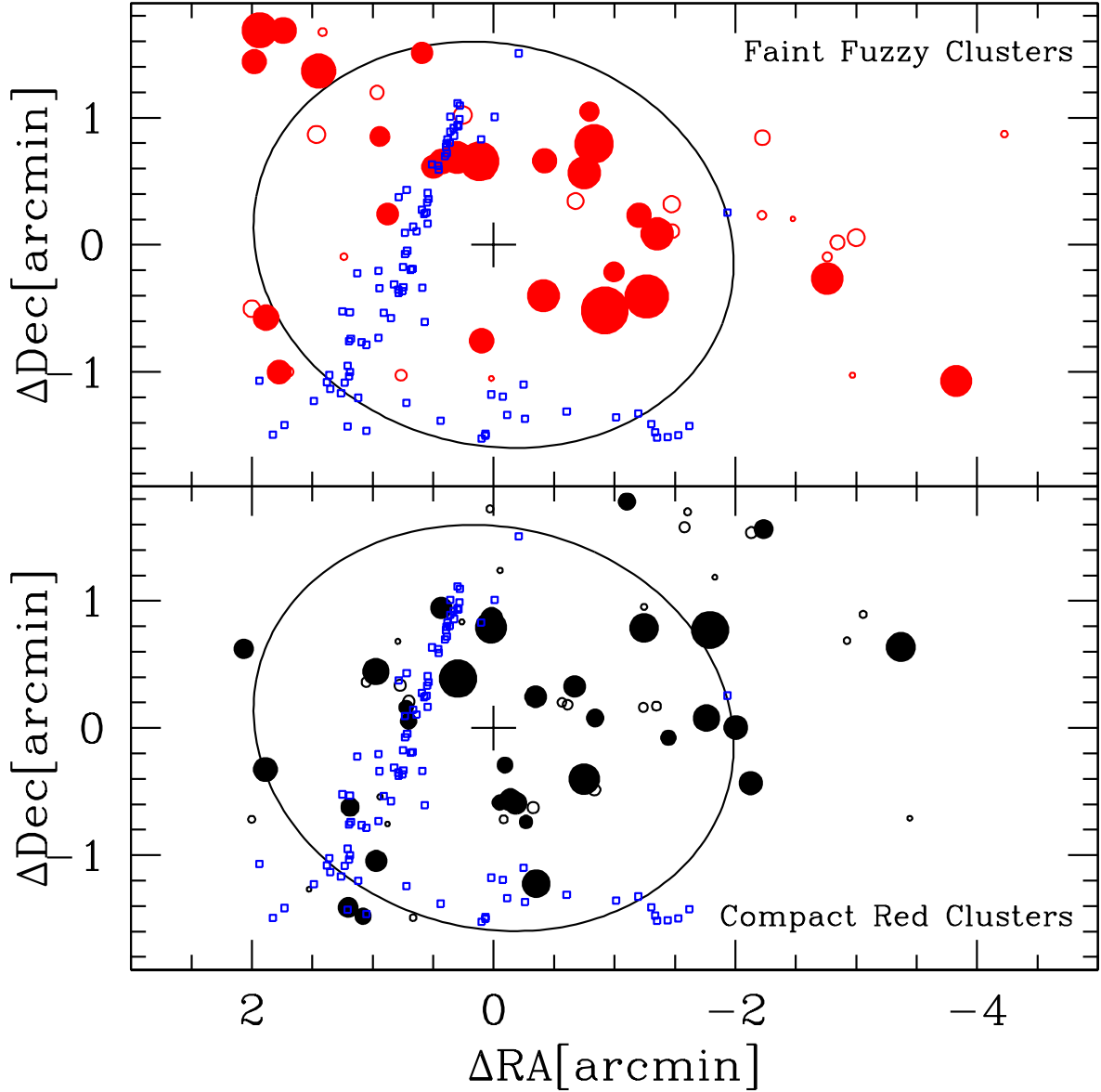}
\caption{Spatial distribution of NGC 5195 star clusters. 
Red circles in the upper panel are faint fuzzy
 clusters, whereas black circles in the lower panel are compact red clusters with $1.0<(V-I)<1.5$. 
 Bright clusters with $V<23.3$ are marked by filled circles and faint clusters with $V \geq 23.3$ 
 by open circles in each panel.
 The size of the circles is proportional to the luminosity of the corresponding clusters:
 the bigger, the brighter.
Squares represent blue clusters with $(V-I)<0.7$. 
 A large ellipse in each panel represents an approximate optical extent of NGC 5195
 with position angle $79^{\rm o}$ and ellipticity $e=0.21$ \cite[]{dev91}.
 A large plus sign in each panel indicates the center of NGC 5195.
\label{fig3}}
\end{figure}

\end{document}